\documentclass[onecollarge]{svjour2}       
\smartqed  
\usepackage{graphicx}
\usepackage{caption}
\usepackage{subfigure}
\newcommand{\rhat}{\mbox{$\mathrm {\mathbf{\hat{r}}}$}}

\begin{document}

\title{A fields only version of the Lorentz Force Law: Particles replaced by their fields}


\author{Philip H. Butler         \and
Niels G. Gresnigt				\and\\
        Martin B. van der Mark		\and
        Peter F. Renaud
}


\institute{Philip H. Butler \at
              Department of Physics and Astronomy\\
							University of Canterbury,\\
							Private Bag 4800, Christchurch, 8140,\\
							New Zealand \\
              \email{phil.butler@canterbury.ac.nz}           
           \and
Niels G. Gresnigt \at
		Department of Physics and Astronomy\\
							University of Canterbury,\\
							Private Bag 4800, Christchurch, 8140,\\
							New Zealand \\
              \email{niels.gresnigt@gmail.com}
           \and
           Martin B. van der Mark \at
		Philips Research Europe,\\
		High Tech Campus 34, WB-21,\\
		5656 AE, Eindhoven,\\
		Netherlands\\
			\email{martin.van.der.mark@philips.com}
           \and
           Peter F. Renaud	\at
		Department of Mathematics and Statistics\\
							University of Canterbury,\\
							Private Bag 4800, Christchurch, 8140,\\
							New Zealand \\
           \email{peter.renaud@canterbury.ac.nz}
}

\date{Received: \today / Accepted: date}

\maketitle

\begin{abstract}
We show that the Lorentz force law, $\mathbf{F}^L_1=q_1(\mathbf{E}+\mathbf{v_1}\times \mathbf{B})$ being the charge on particle $1$ interacting with the electromagnetic fields due to all other particles, can be written in a pure field form $\mathbf{F}^L_1=-\nabla_1 U^{EM}$. In this expression $U^{EM}$ is the total electromagnetic energy of the system of particle $1$ and all other particles.
In deriving this result we review the old but not widely known results that Maxwell's equations follow uniquely from Special Relativity, and that the Lorentz force law follows from applying Hamilton's variational principle to this result.

For a two particle system, the standard view is that the electromagnetic force on particle $1$ is the result of the charge of particle $1$ interacting locally with the field of particle $2$, and conversely. Both charges $1$ and $2$, and fields $1$ and $2$ are needed. In our approach, the fields of all particles contribute to the electromagnetic interaction everywhere, over all of space. The charges of the particles do not enter the theory except incidentally, via GaussÕs law. This has novel interpretational consequences.
In particular, it allows a charged particle to be replaced by its electric and magnetic fields, much as a particle in quantum mechanics is replaced by its complex valued wavefunction.

\end{abstract}

\section{Introduction}
\label{sec:intro}

Lorentz derived his Lorentz transformation laws from the invariance properties of Maxwell's electromagnetic equations. This Special Relativity (SR) was widely known but not widely accepted until Einstein famously derived it from simple assumptions of certain invariances of space, time and the speed of light. Today it is well known, if not widely known, that all of Maxwell's electromagnetism (EM) follows from applying further invariance principles to special relativity. In particular it is often stated that the Lorentz force law is a \textsl{separate yet essential supplement} to Maxwell's equations (see for example page 2 of Jackson \cite{jackson1975cee} or page 782 of Ohanian \cite{ohanian1989pe}). We deduce therefore that is not widely known, although it is well known to some, that by supplementing Maxwell's equations by Hamilton's variational principle, the electromagnetic force laws follow \textsl{also} from this additional basic principle (see page 473 and 475 of Misner, Thorne and Wheeler \cite{misner1973g}).

In this paper we start from SR (Einstein) and summarise Doughty's \cite{doughty1990li} use of the invariance of SR (and essentially nothing else) to get Maxwell's EM.  That is, we extend Einstein's arguments to get the laws of Coulomb, Faraday, Maxwell, Lenz, et al, from symmetry and invariance considerations, and not from experiment (but the result is of course fully in concordance with experiment).
We then show that the Lorentz force law can be derived from the energy density of the electric and magnetic fields (which satisfy Maxwell's equations), together with considerations of the global conservation of energy. That is, only energy conservation is needed to be added to Doughty's arguments to get the Lorentz force.

Although much of our derivation can be found scattered in the literature (see for example section 1.11 of Jackson \cite{jackson1975cee} and in General Relativistic language in\cite{misner1973g}), our approach enables us to take a further step, one of interpretation. The new interpretation is a change to the relationship of electric charges and electric currents to electric and magnetic fields.
Specifically we have a duality between charges and currents and electric and magnetic fields. The duality is similar to the duality between quantum particles and their wave functions.

In section \ref{sec:history} we give our perspective on some key aspects of the standard, historical, approach to deriving the Lorentz force law.
In section \ref{sec:emfield} we give the elements of a Lorentz invariant derivation of Maxwell's equations from SR.
In section \ref{sec:dipole}, rather than following Misner, Thorne and Wheeler \cite{misner1973g} who use the apparatus of General Relativity to derive the Lorentz force law,  we consider the electrostatic situation of the fields due to two  charged spheres, $q_1$ at $\mathbf r_1$, and $q_2$ at $\mathbf r_2$.
In brief, the argument is that electrostatic energy density $u(\mathbf r)$ of the combined field at a point $\mathbf r$, $\mathbf E_1(\mathbf r) + \mathbf E_2(\mathbf r)$, is integrated over all space to give the total electrostatic energy of the system $U^{E}$.
Then, by Hamilton's principle, the change to this total energy when one charge moves, gives the force that acts on that charge.

In more detail, the energy density is
\begin{eqnarray}
u(\mathbf r) & = & \frac{1}{2}\epsilon_0(\mathbf E_1(\mathbf r) + \mathbf E_2(\mathbf r))^2 \nonumber \\
& = & \frac{1}{2}\epsilon_0(\mathbf E_1(\mathbf r)^2 + \mathbf E_2(\mathbf r)^2) + \epsilon_0\mathbf E_1(\mathbf r)\cdot\mathbf E_2(\mathbf r)
\label{eq:ur}
\end{eqnarray}
and the total electrostatic energy is
\begin{eqnarray}
U^{E} & = & \int_{\textrm{all space}} u(\textbf r)\,\textrm {dV}
\label{eq:intallspace}
\end{eqnarray}
Observe that the value of $U^{E}$ changes whenever the location of either of the two charges changes, but only as a result of the term $\mathbf E_1(\mathbf r)\cdot\mathbf E_2(\mathbf r)$.

The electrostatic or Coulomb force, $\mathbf{F}^{C}$, due to the \textsl{field} of $q_2$ on the \textsl{field} of $q_1$ is the ratio of the energy change to the position change
\begin{eqnarray}
\mathbf{F}^{C} & = & -\lim_{\delta r\rightarrow 0} \frac{\delta U^{E}}{\delta r}\rhat \nonumber \\	
& = & -\nabla U^{E}
\label{eq:gradU}
\end{eqnarray}
In the Lorentz force law as usually expressed, the fields act on charges. However in our approach the forces arise directly from changes to the total energy in the total EM field.

Section \ref{sec:dipole} shows this explicitly for the electrostatic situation above. We will evaluate the integral (\ref{eq:intallspace}) for two charged spheres, and then differentiate it with respect to $\delta\mathbf r$ to obtain the Coulomb force law. From Coulomb's force, $\mathbf{F}^{C}$, we may obtain the Lorentz force, $\mathbf{F}^{L}$, via Lorentz transformation to another inertial frame (a relativistic boost).
The reader can carry out similar steps to calculate the force on a test charge inside a parallel plate capacitor, and for the force between two current carrying wires. Some interesting pedagogical issues arise in these familiar situations.


\section{History and perspective}
\label{sec:history}

In terms of the underlying conceptual ideas, Coulomb's law, as with Newton's law of gravity, was historically expressed in terms of {\sl action at a distance} of one charge acting directly on another charge
\begin{eqnarray}
\mathbf F_{12}^{C} =  \frac{1}{4\pi\epsilon_0} \frac{q_1q_2}{r_{12}^2} \rhat_{12}
\label{eq:coul}
\end{eqnarray}
where the relative separation vector between the charges is
\begin{eqnarray}
\mathbf r_{12}=\mathbf r_2 - \mathbf r_1
\end{eqnarray}
and we write $\textbf r_{12} = r_{12}\rhat_{12}$ where $r_{12}$ is the magnitude of the vector $\textbf r_{12}$ and $\rhat_{12}$ its direction.

Equation (\ref{eq:coul}) describes mathematically the experimental observations made by Coulomb in the 1780's, that there exists a force between two static charges separated in space and that the magnitude of this electrostatic force is directly proportional to the magnitude of each charge and inversely proportional to the square of the distance separating the two charges. Coulomb's force law, eq(\ref{eq:coul}), treats the two charges symmetrically, however the action at a distance aspect of the law is contrary to the precepts of Lorentz relativity and so this formulation can only be used in electrostatics.

To visualize the mechanics of the magnetic force between two bodies, in 1852  Faraday introduced {\sl lines of force} \cite{faraday1852pcl}. When iron filings are spread over paper and brought near a bar magnet, the iron filings orient themselves end to end in lines from one pole of the magnet to the other. Faraday interpreted these lines as being the lines of force. Faraday also showed experimentally that these lines of force do not fit action at a distance models \cite{faraday1852pcl}. The lines of force were modified by Maxwell to tubes of force. This modification allowed Maxwell to make fluidic assumptions about the force and to derive a mathematical theory of electromagnetic fields. Maxwell believed these tubes of force propagated through the ether, creating a tension between bodies that was the electromagnetic force \cite{maxwell1865dte,maxwell1856fsl}.

The concept of fields acting as intermediaries allows an object to act on a distant object, and allows the incorporation of retarded fields to allow for the travel time of the information, resolving this problem with Coulomb's force law, eq(\ref{eq:coul}).
The electric field $\mathbf{E}_1$ at point $\mathbf r_2$ generated by the charge $q_1$ at $\mathbf r_1$  is equal to
\begin{eqnarray}
\mathbf{E}_1=\frac{q_1}{4\pi\epsilon_0 r_{12}^2}\mathbf{\hat{r}}_{12}
\end{eqnarray}
The introduction of this field concept in the nineteenth century thus allowed Coulomb's law to be replaced by a new law which describes the {\sl local interaction} of the field due to charge $ q_1$, on charge $q_2$,
\begin{eqnarray}
\mathbf F_{12}^{E} &=&  \mathbf E_1q_2
\label{eq:lore}
\end{eqnarray}
Likewise the effect the field due to  charge $q_2$ on charge $q_1$ is
\begin{eqnarray}
\mathbf F_{21}^{E} &=&  \mathbf E_2q_1
\label{eq:lore21}
\end{eqnarray}
These two equations satisfy Newton's third law, since $\mathbf F_{12}^{E} = -\mathbf F_{21}^{E}$.
We say the charge $q_1$ interacts with the electric field $\mathbf{E}_2$, and also that the field $\mathbf{E}_2$ acts on the charge $q_1$ to give the force law of eq(\ref{eq:lore21}). This is often called the electrostatic Lorentz force law.  Despite the name, it first appeared in a paper by Maxwell in 1861 \cite{clerkmaxwell1861plf}. Three years later, in 1864, Maxwell has this force law as one of his original eight electromagnetic equations \cite{maxwell1865dte}.

The force laws, eq(\ref{eq:lore}) and eq(\ref{eq:lore21}), involve the action of a field (due to one charge) on the other charge, and are thereby unsymmetrical. However, together with the use of retarded fields, this ``Maxwell--Lorentz'' electrostatic force law not only resolved the ``instantaneous action at a distance'' issue of Coulomb's law, it also included the principle of superposition. The principle of superposition is part of the definition of a vector field, and is thus intrinsic to all of Maxwell's equations. As with Coulomb's law, the force experienced by one charge due to a static discrete distribution of other charges may be calculated using the principle of superposition, either by adding the force vectors or  by adding the field vectors.

Given the absence of magnetic monopoles, lifting the restriction that the charges be stationary with respect to one another introduces magnetic fields. There is no law equivalent to Coulomb's law for magnetism and so action at a distance is not an issue that arises. The magnetic force on a charge $q$ is calculated using the magnetic Lorentz force law
\begin{eqnarray}\label{eq:lorb}
\mathbf{F}^{M}_{12}=q_1(\mathbf{v}_1\times \mathbf{B}_2)
\end{eqnarray}
where $\mathbf{B}_2$ is the magnetic field produced from charge or charges $q_2$  moving at some velocities $\mathbf{v}_2$ with respect to the laboratory frame.

The magnetic field at a point $\mathbf r_1$ can be calculated using the Biot-Savart law
\begin{eqnarray}
d\mathbf{B}_2(\mathbf r_1)=\frac{\mu_0}{4\pi}\;\frac{i_2d\mathbf{s}_2\times \mathbf{\hat{r}}_{12}}{r_{12}^2}
\label{eq:biot}
\end{eqnarray}
relating an infinitesimal magnetic field $d\mathbf{B}_2$ at $\mathbf{r}_1$ due to a infinitesimal current element $i_2d\mathbf{s}_2$ at  $\mathbf r_2$.

The general Lorentz force law, which incorporates both electric and magnetic fields, is the sum of forces (\ref{eq:lore}) and (\ref{eq:lorb})
\begin{eqnarray}\label{eq:lorentz}
\mathbf{F}^L=q_1(\mathbf{E}_2+\mathbf{v}_1\times \mathbf{B}_2)
\end{eqnarray}
This law may be derived from the electrostatic force (\ref{eq:lore}) in a frame where the magnetic field is zero, to the frame with the non-zero magnetic field, by means of a Lorentz boost. The Lorentz force law provides an electrodynamic theory of charges where both the electric and magnetic fields are mediators for electromagnetic force and both fields act on the charge $q_1$.

Historically, the Faraday-Maxwell approach led to Maxwell's equations, and then Lorentz, Poincar\'{e} and others found the invariance properties of the equations were not those of Galileo and Newton, but involved time in a new way. In 1905 Einstein reversed this process and showed that these Lorentz and Poincar\'{e} transformations followed from extending the Galilean spatial transformations to spacetime, that is special relativity (SR), by assuming the invariance of the speed of light. In the next section we review the argument that Maxwell's equations follow from SR.


\section{A Lorentz Invariant Derivation of Maxwell's Equations}\label{sec:emfield}

The above historical approach to deriving the Lorentz force law is usually the approach taken in undergraduate texts on electromagnetism. In this approach the Lorentz force law (\ref{eq:lorentz}) (or its non-relativistic limit, (\ref{eq:lore})) is also used to derive Maxwell's equations. Several modern authors give a different, Lorentz invariant approach to Maxwell's equations that does not use the Lorentz force law. One such derivation is provided in chapter 18 of Doughty \cite{doughty1990li}. The argument is, in outline:

First use Einstein's logic to deduce special relativity from the homogeneity of space and time, isotropy of space, and the invariance of the speed of light. Second ask for the simplest non-trivial vector field $A_\mu$, such that $A_\mu A^{\mu}$ is a Lorentz scalar. The derivative (or 4-curl) of this field is a second rank anti-symmetric tensor called the electromagnetic (or Faraday) tensor and written $F^{\mu\nu}$
\begin{eqnarray}
F^{\mu\nu}=\partial^{\mu}A^{\nu}-\partial^{\nu}A^{\mu}
\end{eqnarray}
The electric and magnetic field components  are contained in this tensor, $E^i=F^{0i}$ and $B^i=F^{jk}$.

Maxwell's field equations can be written as the derivative of this tensor and its dual
\begin{eqnarray}
\partial_\nu F^{\mu\nu}=\mu_0J^{\mu}\quad\mathrm{and}\quad \partial_{\nu}\tilde{F}^{\mu\nu}=0 \label{eq:maxwell}
\end{eqnarray}
where the source $J^{\mu}$ is another 4-vector, the electromagnetic current density.

Next, construct the energy-momentum tensor $T^{\mu\nu}$ of the electric field. Because energy-momentum is a physical quantity that can be measured, it cannot have any gauge-freedom and so the energy-momentum tensor is constructed in terms of the gauge-invariant tensor $F^{\mu\nu}$. Quoting Doughty: \textit{``$T^{\mu\nu}$ must be a proper tensor symmetric with respect to $\mu$ and $\nu$. No symmetric tensor may be constructed from terms having only a single factor of $F^{\mu\nu}$ combined with the universal tensors $\eta_{\mu\nu}$ and $\epsilon^{\mu\nu\lambda\rho}$. Noting also that $T^{\mu\nu}$ for a particle is quadratic in the time derivatives $\dot{z}^\mu$ of the dynamical variables, we consider quadratic combinations of $F^{\mu\nu}$ which itself contains spacetime derivatives of $A^{\mu}$"}.

Because the indices $\mu$ and $\nu$ cannot both be on a single factor of $F^{\mu\nu}$ in any term of $T^{\mu\nu}$, there are only two independent terms that may be used \footnote{To be precise, there are a total of four second-rank tensors quadratic in $F^{\mu\nu}$. However one of these is not proper and another is a linear combination of the other two proper tensors.}. The first of these is a term of the form $F^{\mu\lambda}F^\nu_\lambda$ which is symmetric in $\mu\nu$. The second term is obtained by instead placing the indices on a factor of $\eta^{\mu\nu}=\mathrm{diag}(1,-1,-1,-1)$ the Minkowski metric which is also symmetric. One must then multiply by the invariant quadratic $F_{\lambda\pi}F^{\lambda\pi}$ to ensure the there is common dimensionality.

The electromagnetic energy-momentum tensor must therefore be of the form,
\begin{eqnarray}
T^{\mu\nu}=aF^{\mu\lambda}F^\nu _\lambda+b\eta^{\nu\mu}F_{\lambda\pi}F^{\lambda\pi}
\end{eqnarray}
where $a$ and $b$ are constants.

By demanding that this tensor be conserved identically as a result of the free field Maxwell equations and by coupling the electromagnetic field to any system for which the energy details are known, the constants $a$ and $b$ may be solved for. One finds that $a=1/\mu_0$ and $b=-a/4$. The electromagnetic energy-momentum tensor is therefore \footnote{A more rigorous derivation of the special relativistic electromagnetic energy-momentum tensor using Noether's theorem can be found in chapter 19 of Doughty \cite{doughty1990li}.}:
\begin{eqnarray}
T^{\mu\nu}=\frac{1}{\mu_0}\left(F^{\mu\lambda}F^{\nu}_{\lambda}-\frac{1}{4}\eta^{\mu\nu}F_{\lambda\pi}F^{\lambda\pi}\right)
\end{eqnarray}
The energy density is then recognised as the $T^{00}$ component of the electromagnetic energy momentum tensor
\begin{eqnarray}
T^{00}=\frac{1}{2}(\epsilon_0\mathbf{E}^2+\frac{1}{\mu_0}\mathbf{B}^2)
\end{eqnarray}

This argument allows Doughty and others to derive Maxwell's equations from special relativity. This reverses the historical approach which starts with Coulomb's law, and by a series of generalisations and other extensions, arrives at Maxwell's equations and their invariance properties, the Lorentz transformation laws. Thus Maxwell's equations imply special relativity, and vice versa.

\section{Coulomb's Law From Hamilton's Principle \label{sec:dipole}}

In this section we show that by considering the fields generated by the charges, and not the charges themselves,  we are able to express the electrostatic force symmetrically in terms of retarded fields.
We evaluate the integral (\ref{eq:intallspace}) for two charges, and then differentiate it with respect to $\delta\mathbf  r_{12}$.

The electric fields associated with the two charges $q_1$ and $q_2$ are given by Gauss's law, a special case of Maxwell's equations, eq(\ref{eq:maxwell})
\begin{eqnarray}
\mathbf{E}_1(\mathbf r)=\frac{q_1}{4\pi\epsilon_0 r^2_1}\mathbf{\hat{r}}_1,\qquad \mathbf{E}_2(\mathbf r)=\frac{q_2}{4\pi\epsilon_0 r^2_2}\mathbf{\hat{r}}_2,
\end{eqnarray}
where $\mathbf{r}_1$ and $\mathbf{r}_2$ are the vectors from $q_1$ and $q_2$ respectively to a given point ${\mathbf r}$ in space where the electric fields are measured. The vectors $\mathbf{\hat{r}}_1$ and $\mathbf{\hat{r}}_2$ point radially outward from the charges.

\begin{figure*}[ht]
  \centering
		\includegraphics[width=0.70\textwidth]{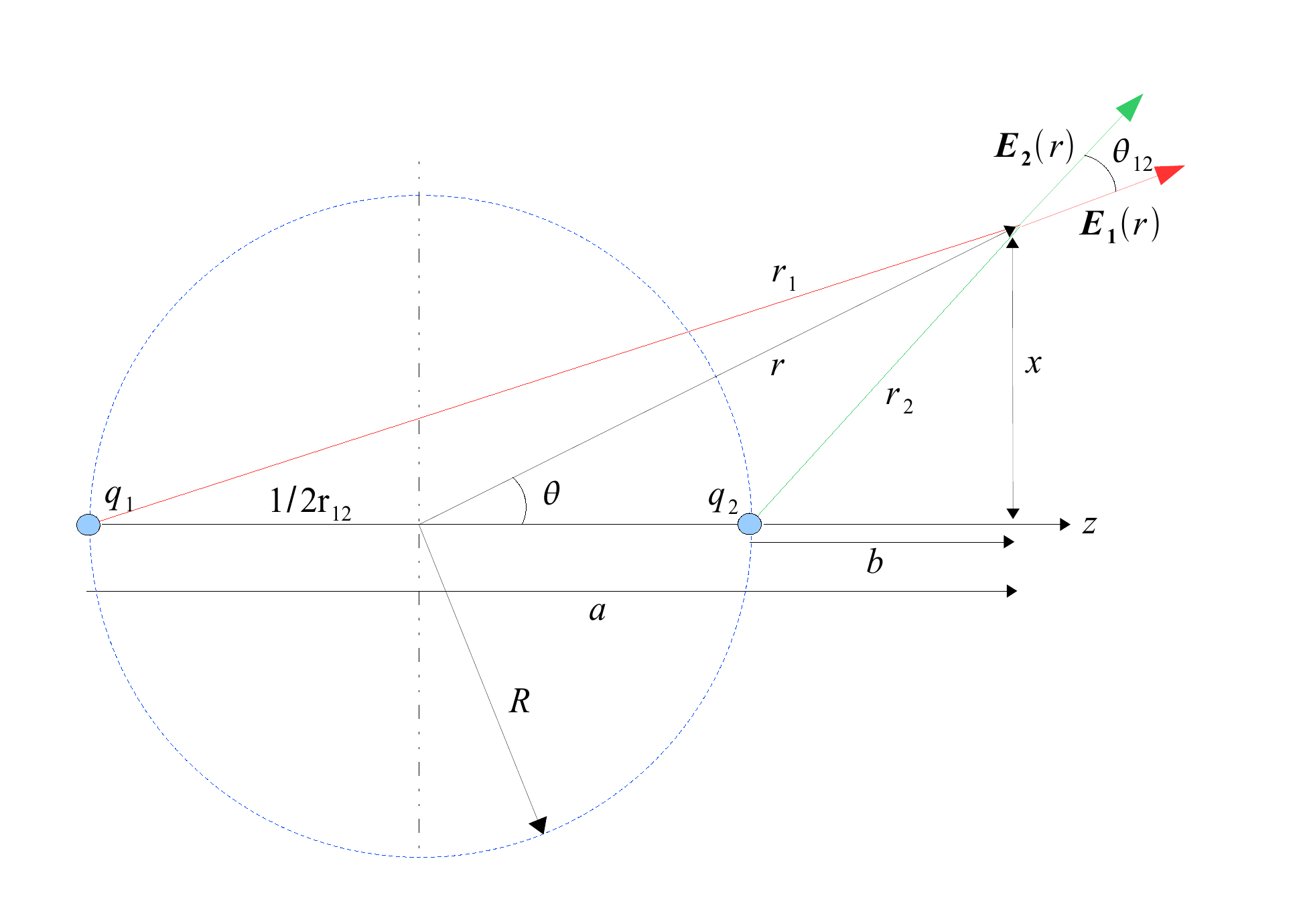}
	\caption{Two charges $q_1$ and $q_2$ separated in space by $r_{12}$. The energy density at any point inside the circle is negative. Outside the circle it is positive. }
	\label{fig:twocharges1}
\end{figure*}

A pictorial representation of the two charge system is given in Figure \ref{fig:twocharges1}. In the spherical coordinates $(r,\theta,\phi)$ of the figure, $q_1$ is at $(R,\pi,0)$ and $q_2$ at $(R,0,0)$.

The energy density of the system is
\begin{eqnarray}
u_{12}(\mathbf r)&=&\frac{\epsilon_0}{2}(\mathbf{E}_1(\mathbf r)+\mathbf{E}_2(\mathbf r))^2 \nonumber \\
&=& \frac{\epsilon_0}{2}{E}_1(\mathbf r)^2+\frac{\epsilon_0}{2}{E}_2(\mathbf r)^2+\epsilon_0{E}_1(\mathbf r) {E}_2(\mathbf r) \cos\theta_{12}
\end{eqnarray}
where $\theta_{12}$ is the angle from $\mathbf{\hat{r}}_1$ to $\mathbf{\hat{r}}_2$.

The integrals over all space of the first two terms,  $u_1=\frac{1}{2}\epsilon_0 {E}_1^2$ and $u_2=\frac{1}{2}\epsilon_0 {E}_2^2$, are independent of the positions of $q_1$ and $q_2$, and so the $u_1$ and $u_2$ terms can be ignored in the calculations of the force. It is clear that these terms describe the self-interactions. We assume that the charged objects have a finite radius, so the integrals are finite and constant.

The interaction energy density term is
\begin{eqnarray}
u_{1,2}&=&  \epsilon_0 \mathbf{E}_1 \cdot \mathbf{E}_2 \nonumber \\
	&=& \epsilon_0 {E}_1{E}_2\cos\theta_{12}
\end{eqnarray}

Substituting the expressions for ${E}_1$ and ${E}_2$ and using some trigonometry, see Figure \ref{fig:twocharges1}, one finds after some relatively straightforward calculations that the total interaction energy for the two charge system is given by
\begin{eqnarray}
U_{1,2}=\frac{q_1q_2}{4\pi\epsilon_0}\frac{1}{r_{12}}
\end{eqnarray}
and the force between the two charges is
\begin{eqnarray}
\mathbf{F}_{12}&=&-\nabla_{12}U_{1,2}  \nonumber  \\
     &=&-\frac{\partial}{\partial r_{12}}U_{1,2} \, \rhat_{12}  \nonumber  \\
     &=&\frac{q_1q_2}{4\pi\epsilon_0 r_{12}^2}\rhat_{12}
\end{eqnarray}
which is Coulomb's law.

There are some notational subtleties in the expression above, in that $\mathbf{F}_{12}$ is the force on $q_2$ due to $q_1$,  $\nabla_{12}U_{1,2}$ denotes the change in $U_{1,2}$ as a function of variations in $\mathbf r_{12}$, $U_{1,2}$ is the total interaction energy in the electric field due to the system of $q_1$ plus $q_2$, and  $\mathbf{r}_{12}$ is the vector from $q_1$ to $q_2$.
If we hold a charged sphere $q_1$ stationary at $\mathbf r_1$, and move a second charged sphere $q_2$ at $\mathbf r_2$ by a distance  $\delta\mathbf r_2$, then $\delta\mathbf  r_{12} = \delta\mathbf  r_{2}$.

An interesting observation is that the interaction energy $U_{1,2}$ is negative inside the sphere of radius $R=\frac{1}{2}r_{12}$ centred  midway between the two charges: $U_{1,2} = \frac{1}{2R}-\frac{\pi}{4R}$.  The  interaction energy inside this sphere contributes an attractive component to the force between two like charges. This volume contributes around $-23\%$ to the total energy. On the boundary of this sphere the interaction energy density is zero because the fields are orthogonal. Outside the sphere the interaction energy density is  positive, giving rise to the overall repulsive force between like charges. The volume outside radius $R$ contributes about $123\%$.

\begin{figure}[ht]
  \begin{center}
    \subfigure[The interaction energy in the volume between the sphere of approximate radius $1.62R$ and the sphere of radius $R$ is positive and cancels the negative interaction energy of the smaller sphere.] {\label{fig:2a}\includegraphics[scale=0.45]{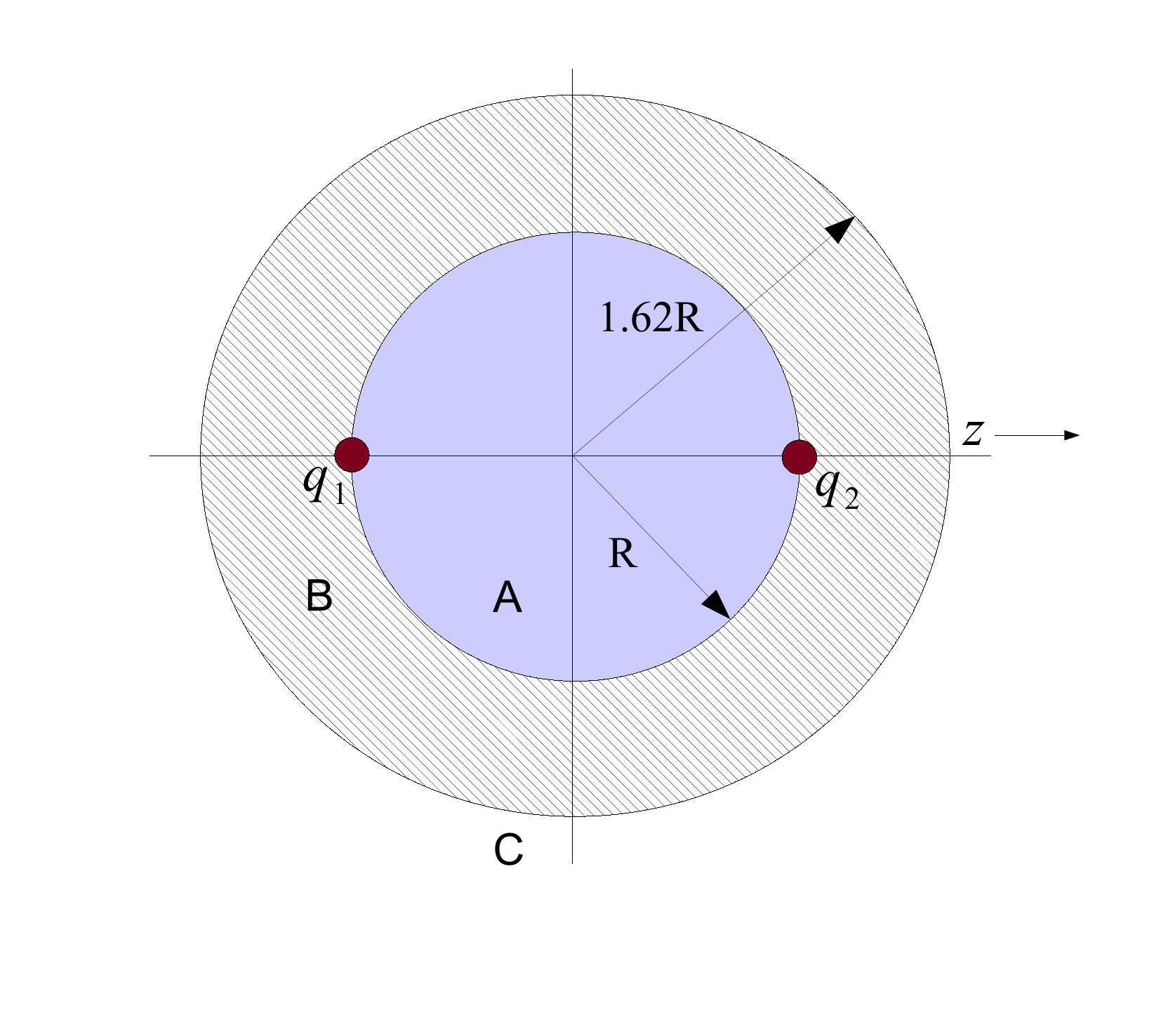}}
    \subfigure[The contribution from the volume between the two charges (the part of space between $z=-R$ and $z=R$) is zero because the contribution  for cylindrical coordinate $r$ in the range from $0$ to the surface of a sphere of radius $R$, cancels (for each value of $z$) the contribution outside the sphere.] {\label{fig:2b}\includegraphics[scale=0.45]{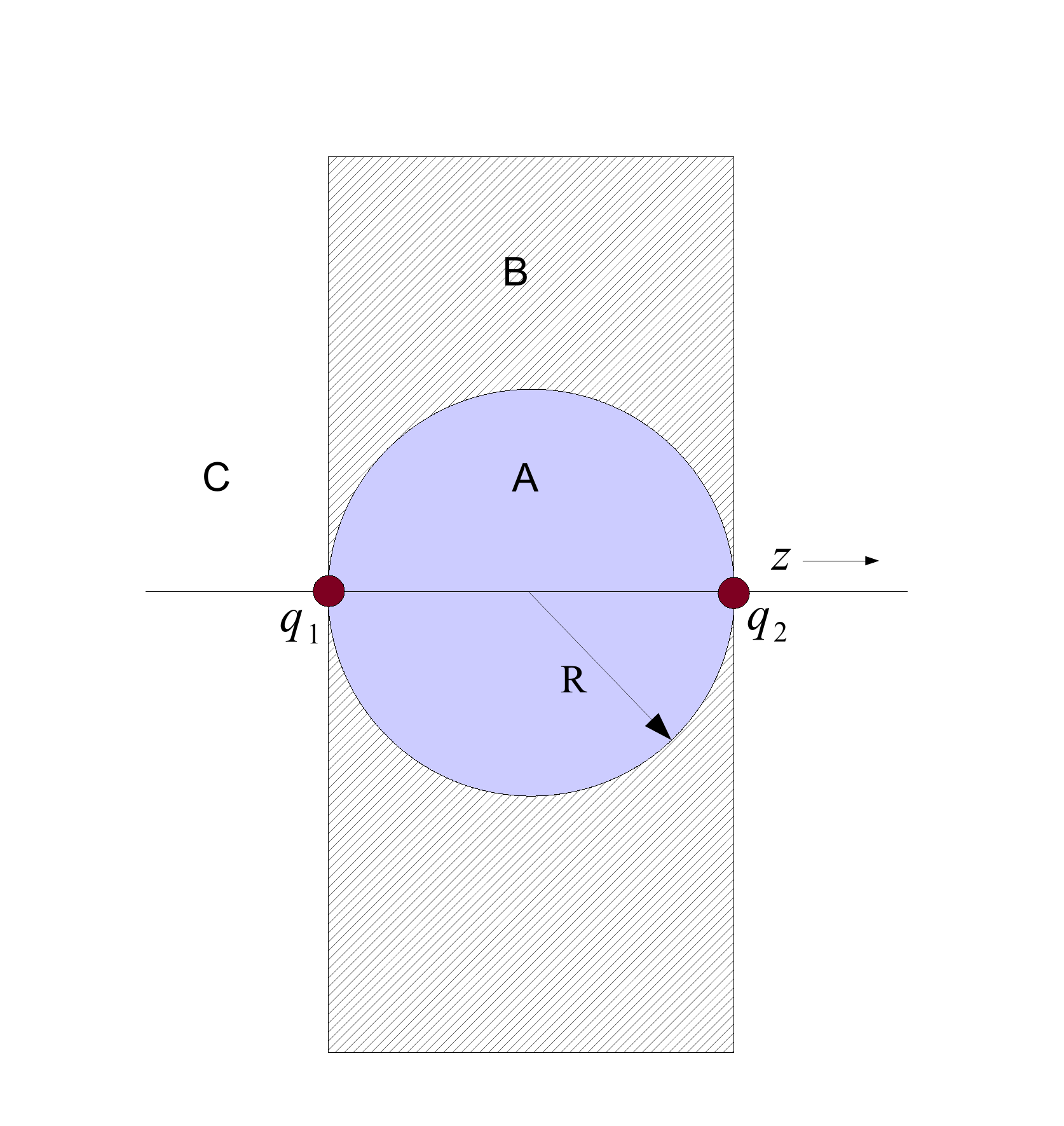}}
    \subfigure[The negative interaction energy inside the sphere of radius $R$ is  negated by another sphere of radius $2R$ centered on either of the two charges.] {\label{fig:2c}\includegraphics[scale=0.45]{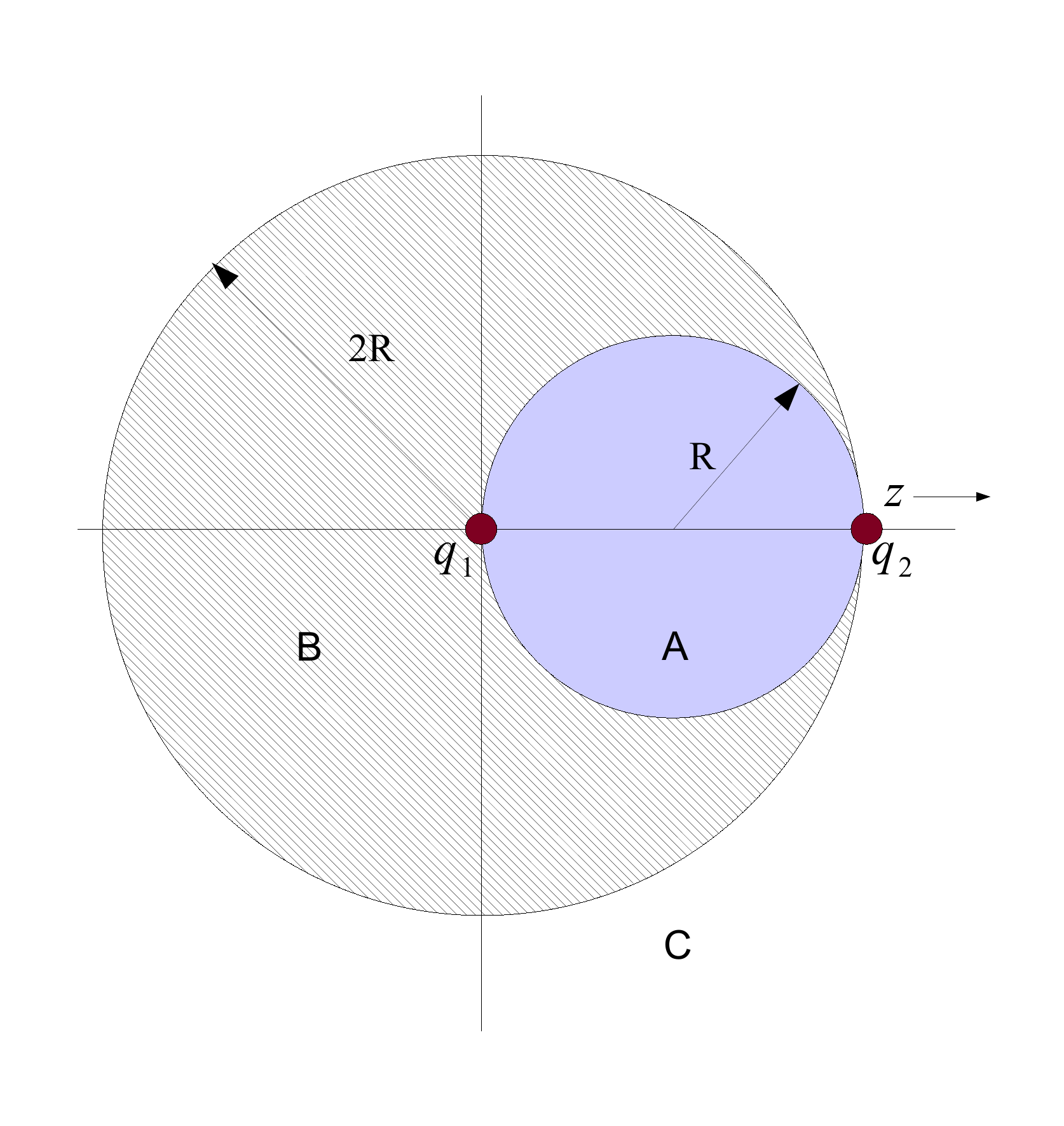}}
  \end{center}
  \caption{This figure shows three regions of space (each containing the sphere of radius $R$ of negative interaction energy) within which the total interaction energy is zero.}
  \label{fig:99}
\end{figure}

There exist several noteworthy regions of space within which the total interaction energy is zero. One of these is found by asking: what must the radius $R_0$ of a sphere concentric with the aforementioned sphere of radius $R$ be so that the interaction energy inside this sphere vanishes? See Figure \ref{fig:2a}. A straightforward calculation reveals that $R_0$ is approximately $1.62R$.

Secondly, the net field contribution is zero from the infinite slab between the charges, that is the region of space where $-R\leq z\leq R$. Indeed using cylindrical coordinates, it is easy to show that for each value of $z$ in this range, the negative contribution to $\int f(r,\theta)\,r^2\sin\theta\,dr$ from inside the sphere of radius $R=\frac{1}{2}r_{12}$ centered at the coordinate origin, cancels the contribution from $R$ to $\infty$, see Figure \ref{fig:2b}.

Thirdly, the negative interaction energy inside the sphere of radius $R$ is also cancelled by another sphere of radius $2R$ centered on either of the two charges, as shown in Figure \ref{fig:2c}. This result follows by evaluating the integral eq(\ref{eq:intallspace}) for a sphere centered on $q_1$.
Alternatively, this can be seen from the fact that the field of point charge $q_1$ for $r>R$, is the same as that of a uniform shell of radius $R$ with total charge $q_1$. The shell has zero field inside $r$, and therefore both shell and point charge have zero interaction energy with $q_2$ in the volume inside $r$, as long as $r<2R$.

As an explicit demonstration of our claim that the  Coulomb force law follows from the EM energy density and Hamilton's variational principle, we have evaluated the force between two charges at rest. It is not difficult to repeat the point charge calculation for a test charge in a capacitor, and the force between a pair of current-carrying wires, but each example carries some subtleties in choosing coordinate systems and the limits of the respective integrals. Boosting any of these results to a moving frame using the Lorentz transformation laws, which are the transformation laws for Maxwell's equations, leads to the full, relativistic Lorentz force law for the interactions between moving electric charges and for interactions between electric currents.

\section{Discussion}\label{discussion}

We have reviewed the old but not well known result that Maxwell's equations follow from special relativity.  By supplementing Maxwell's equations by Hamilton's variational principle that force is related to the gradient of energy, $\mathbf F = - \nabla U$, we have shown that the Lorentz force law is contained within both Maxwell's equations and special relativity, a result not quite so old, and even less widely known.

Many authors, and essentially all undergraduate texts, follow the historical development and deduce Maxwell's equations by a series of extensions to Coulomb's force law, the first extension being Faraday's concept of lines of force. These lines of force were introduced as acting on charges and currents, but Maxwell recognised that the fields have merit in themselves, as is reflected in the homogeneous Maxwell equations which contain only the fields, and give rise to electromagnetic waves.

In the standard formulation, the Lorentz force law eq(\ref{eq:lorentz}) has electric and magnetic fields as the mediators of electromagnetic force and it gives the electromagnetic force on a (perhaps moving) test charge through the electric and magnetic fields at the position of the test charge. This means that the Lorentz force law is essentially local, what happens in the rest of space does not affect the force experienced by the test charge. However the Lorentz force law is not symmetric with respect to the charge (or charges) that are the source of the field, and the charge that is being acted upon by the field.

Our expression for the Lorentz force law is new, it is in terms of the interactions of fields on fields, not fields on charges. Our approach thus requires a re-interpretation of the usual statements that electromagnetic fields do not interact with other electromagnetic fields but only with charges and currents (for an example see the statement on page 226 of Aitchison and Hey \cite{aitchison1982gtp}).

We have shown that working with only the fields, one is able to obtain a force law which is both Lorentz invariant and symmetrical. The fields of the charges are the mediators of the force, but this time not through the interaction of the fields with the test charge at a single point in space, but rather through the interaction of the fields with the fields from the test charge throughout the entire causally connected universe. Although the fields interact at points, and then as the energy density at points, the net force at a time $t$ results from considering all the points in space at that time $t$.
We note that this kind of dispersed interaction seems similar in some aspects to the delocalization of the wave function in quantum mechanics before it collapses at (what is usually supposed to be) the point of interaction.

In our formulation, the Lorentz force law is non-local, in much the same sense that gravity is non-local in general relativity. Consider only the Coulomb part. While the interaction energy density $u_{1,2}=\epsilon_0 \mathbf{E}_1\cdot\mathbf{E}_2$ is local, the Coulomb force comes from $-\nabla_{12}\int{u_{1,2}}dV$. The repulsive force between two positive charges $q_1$ and $q_2$ separated by $r_{12}$ comes not from the integral close to the charges, but from the integral over all space. Indeed, the integral within the sphere of radius $r_{12}/2$ leads to an attractive contribution (of around $23\%$), a contribution that is overwhelmed by the repulsive integral (of $123\%$) of the interaction energy density outside $r_{12}$.

The fields-only approach, eq(\ref{eq:gradU}), proposed here is thus different from conventional EM in two respects: It is symmetric with respect to the sources of the fields;  and the force experienced by each of the sources is determined by the electromagnetic fields due to all charges throughout all space.

There are a number of implications of this approach. There needs to be a re-interpretation of the assumption that electromagnetic fields do not interact with each other. We have shown in this paper that by laying aside this assumption and looking at the quadratic form that is the energy density, one may successfully calculate the electromagnetic forces between charged particles. Our assumption is that fields act on fields, but we emphasise that it is not through their linear superposition properties, but rather through their non-linear, quadratic, contribution to energy density. In our approach, the charges are essentially delocalized, and replaced by their electromagnetic fields in a manner similar to the delocalization of particles in quantum mechanics, where the particles are replaced by their wave functions.

Furthermore in our approach, the electromagnetic property of the charges themselves is reduced to merely being the sources of the fields. Namely, the charges don't interact with each other through electric and magnetic fields, it is the fields themselves that interact through eq(\ref{eq:gradU}): $\mathbf{F}_{12}=-\nabla_{12} U_{12}$.

\begin{acknowledgements}
This work was inspired by a discussion between P.H.B. and M.v.d.M on topological fields and the (inter)action of fields on fields. It is a pleasure to acknowledge the implicit contribution to that conversation by John Williamson \cite{williamson1997}, and for his insightful comments on drafts of this paper. We also acknowledge Jake Gulliksen for implementing some of this work in his B.Sc(Hons) project report \cite{gulliksen2008}.
\end{acknowledgements}

\bibliography{Niels}  
\bibliographystyle{unsrt}


\end{document}